\documentclass[]{spie}  

\usepackage{amsmath,amsfonts,amssymb}
\usepackage{graphicx}
\usepackage[colorlinks=true, allcolors=blue]{hyperref}

\title{CNN-Based Reanalysis of Optical Turbulence at the Canary Islands Observatories}

\author[a]{Airam Salas}
\author[b,c]{Begoña García-Lorenzo}
\author[b,c]{Julio A. Castro-Almazán}
\author[d]{Gelys Trancho}
\affil[a]{Independent Researcher, 38200 La Laguna, Tenerife, Spain}
\affil[b]{Instituto de Astrofísica de Canarias, C/ Vía Láctea s/n, 38205 La Laguna, Tenerife, Spain}
\affil[c]{Universidad de La Laguna, 38200 La Laguna, Tenerife, Spain}
\affil[d]{NSF NOIRLab, Tucson, AZ 85719, USA}
\authorinfo{Further author information: (Send correspondence to B.G-L.)\\B.G-L.: E-mail: begona.garcia@iac.es}

\begin{document} 
\maketitle

\begin{abstract}

Atmospheric optical turbulence, caused by refractive-index fluctuations driven by wind and temperature inhomogeneities, is the primary source of image degradation in ground-based telescopes. Its vertical distribution, described by the refractive-index structure constant (C$_n^2$(h)), determines essential parameters such as seeing ($\epsilon_0$), isoplanatic angle ($\theta_0$), and coherence time ($\tau_0$), which characterize astronomical sites and are also crucial for the design and performance of high-resolution instruments such as adaptive optics. The {\it {Observatorios de Canarias}} (OCAN), comprising the {\it {Observatorio del Roque de los Muchachos}} (ORM) on La Palma and the {\it {Observatorio del Teide}} (OT) on Tenerife, have been extensively characterized through long-term Generalized SCIDAR campaigns. In total, the database comprises nearly 300 observing nights (165 at ORM and 130 at OT) and hundreds of thousands of individual C$_n^2$(h) profiles.

In this work, we present a reanalysis of the OCAN database using deep-learning techniques. An end-to-end software framework was developed to ingest, preprocess, and transform the original turbulence measurements into standardized numerical representations suitable for machine-learning analysis. Temporal–vertical turbulence heatmaps are generated under multiple temporal configurations and enriched with seasonal information. These representations are processed by pre-trained convolutional neural networks (CNN) used as frozen feature extractors to obtain compact deep embeddings.

Unsupervised clustering is subsequently applied to identify latent atmospheric regimes within the dataset. The resulting clusters reveal coherent turbulence structures and offer a complementary perspective on atmospheric conditions at the Canary Islands observatories. This methodology provides new insights into turbulence variability at the Canary Islands sites and represents a novel approach to large-scale turbulence characterization using modern deep-learning techniques.
\end{abstract}


\section{INTRODUCTION}
\label{sec:intro}  

Atmospheric optical turbulence is the primary source of image degradation in ground-based astronomy. It originates from random fluctuations of the atmospheric refractive index driven by temperature inhomogeneities and wind-induced mixing, producing wavefront distortions that limit the angular resolution achievable by terrestrial telescopes. The vertical distribution of turbulence is quantified through the refractive-index structure constant profile, (C$_n^2$(h)), from which the main astroclimatic parameters relevant to adaptive optics (AO) systems are derived, including the seeing ($\epsilon_0$), the isoplanatic angle ($\theta_0$), and the atmospheric coherence time ($\tau_0$). Consequently, accurate characterization of the vertical turbulence structure is essential for astronomical site testing, observatory operations, and the design and optimization of AO systems.

The Generalized SCIDAR technique (e.g. [\citenum{1997Avila}]) is one of the most powerful remote-sensing methods for measuring atmospheric optical turbulence profiles, providing high-resolution estimates of (C$_n^2$(h)) over a wide range of altitudes. Extensive Generalized SCIDAR campaigns have been conducted at major astronomical observatories worldwide, generating large databases that have enabled detailed studies of turbulence and AO-related parameters.

Among these efforts, the {\it{Observatorios de Canarias}} (OCAN), comprising the {\it {Observatorio del Roque de los Muchachos}} (ORM) on La Palma and the {\it{Observatorio del Teide}} (OT) on Tenerife, were the subject of extensive Generalized SCIDAR monitoring campaigns aimed at characterizing the vertical distribution of atmospheric optical turbulence ([{\citenum{2011aGarcia-Lorenzo}, \citenum{2011bGarcia-Lorenzo}]). These campaigns produced a database spanning nearly 300 observing nights in total (165 at ORM and 130 at OT) and containing hundreds of thousands of individual turbulence profiles. Previous analyses of these datasets focused mainly on the statistical characterization of turbulence profiles and derived astroclimatic parameters, revealing common turbulence structures, seasonal behaviour, and overall similarities between the two observatories ([\citenum{2009Garcia-Lorenzo}]).

While traditional studies have relied primarily on integrated atmospheric parameters and statistical profile analyses, the increasing availability of large turbulence databases opens new opportunities for data-driven approaches. Advances in deep learning have demonstrated that convolutional neural networks (CNNs) can extract highly informative representations from complex structured datasets through transfer-learning strategies. When combined with unsupervised learning techniques, these representations can reveal hidden patterns and latent structures that may remain inaccessible to conventional statistical analyses.

The objective of this work is to perform a large-scale reanalysis of the complete OCAN turbulence database using CNN-based feature extraction and unsupervised clustering. Rather than focusing exclusively on integrated AO parameters, we investigate whether distinct atmospheric regimes can be identified directly from the full temporal and vertical structure of turbulence profiles. To achieve this, we developed a complete processing framework that transforms Generalized SCIDAR C$_n^2$(h) profiles into multi-channel turbulence representations suitable for deep-learning analysis. The resulting embeddings are subsequently clustered to explore the existence of recurrent turbulence behaviours and to provide a new perspective on atmospheric variability at the OCAN.

\section{Turbulence Profiles Dataset Description}
\label{sec:dataset}

The dataset analysed in this work originates from the atmospheric turbulence monitoring programmes conducted at the OCAN during the 2000s. These campaigns formed part of a broader effort to characterize the atmospheric conditions of major astronomical sites for future instrumentation and telescope projects, while also supporting long-term studies of atmospheric turbulence and adaptive optics applications at the OCAN.
 
The OCAN comprises the ORM on La Palma and the OT on Tenerife, both located at altitudes close to 2400 m above sea level. Turbulence measurements were obtained using the Generalized SCIDAR technique operating at the 1-m Jacobus Kapteyn Telescope (1m-JKT, ORM) and the 1.5-m Carlos Sánchez Telescope (1.5mTCS, OT). The resulting database comprises nearly 300 observing nights in total (165 at ORM between 2004 and 2009, and 130 at OT between 2002 and 2008), yielding several hundred thousand individual C$_n^2$(h) profiles describing the vertical distribution of atmospheric optical turbulence above these astronomical sites. The data span multiple years and seasons, providing broad coverage of the atmospheric variability at both sites. 

Previous studies based on these observations focused on the statistical characterization of turbulence profiles and derived astroclimatic parameters, including $\epsilon_0$, $\theta_0$, $\tau_0$, and turbulence-layer distributions ([\citenum{2011aGarcia-Lorenzo}, \citenum{2011bGarcia-Lorenzo}, \citenum{2009Garcia-Lorenzo}]). These analyses showed that the two observatories share similar large-scale turbulence structures and exhibit comparable seasonal trends, with most of the optical turbulence concentrated within the boundary layer and recurrent high-altitude turbulent layers appearing during specific periods of the year. Figure \ref{fig:statiscalprofiles} presents the monthly statistical C$_n^2$(h) profiles derived from the OCAN database and provides an overview of the vertical turbulence distributions at each site.

   \begin{figure} [ht]
   \begin{center}
   \begin{tabular}{c} 
   \includegraphics[width=16.5cm]{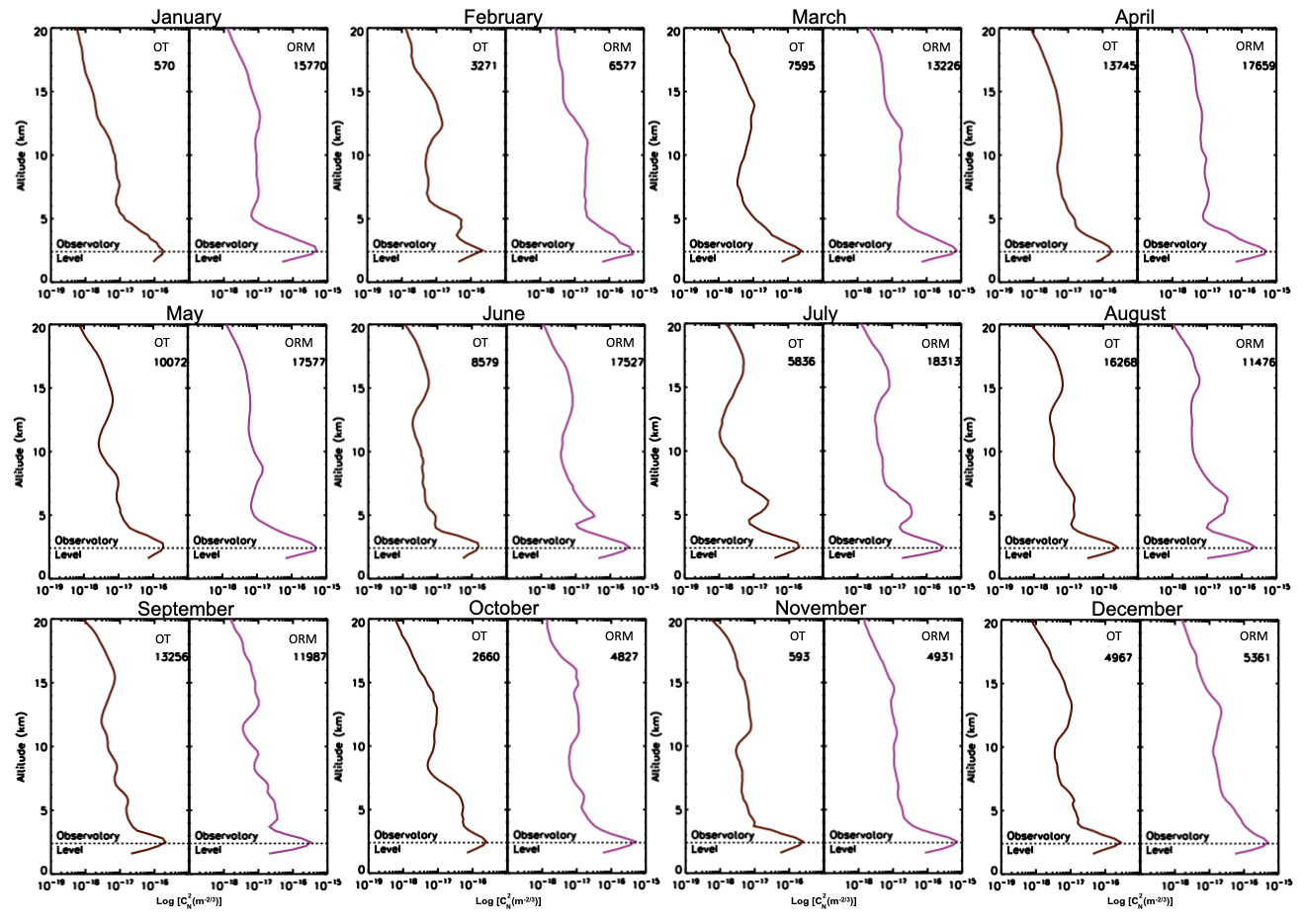}
   \end{tabular}
   \end{center}
   \caption[ ] 
   { \label{fig:statiscalprofiles} 
Monthly statistical C$_n^2$ profiles for the Roque de los Muchachos Observatory (ORM, purple profiles) and the Teide Observatory (OT, brown profiles) derived from Generalized SCIDAR measurements. Profiles were obtained by combining all available turbulence measurements acquired during each month over the full observing campaigns. The horizontal axis shows the turbulence strength, C$_n^2$, in logarithmic scale (m$^ {-2/3}$), while the vertical axis represents altitude above sea level. The dashed horizontal line indicates the observatory altitude ($\sim$2400 m, a.s.l.). The numbers shown in each panel correspond to the total number of individual turbulence profiles contributing to the monthly statistics. Both sites exhibit similar large-scale turbulence structures and seasonal evolution, with most of the turbulence concentrated near the observatory level and recurrent elevated layers appearing at characteristic altitudes throughout the year.} 
   \end{figure} 

\section{DATA PROCESSING PIPELINE}

For the present work, the complete OCAN database was reprocessed and transformed into a unified format suitable for large-scale machine-learning (ML) analysis. Rather than analysing individual turbulence profiles or integrated atmospheric parameters alone, the database is exploited as a continuous temporal record of atmospheric behaviour, enabling the construction of temporal–vertical turbulence representations from which latent turbulence regimes can be investigated through deep-learning techniques. Figure~ \ref{fig:infografia} summarizes the processing workflow.

   \begin{figure} [ht]
   \begin{center}
   \begin{tabular}{c} 
   \includegraphics[width=16cm]{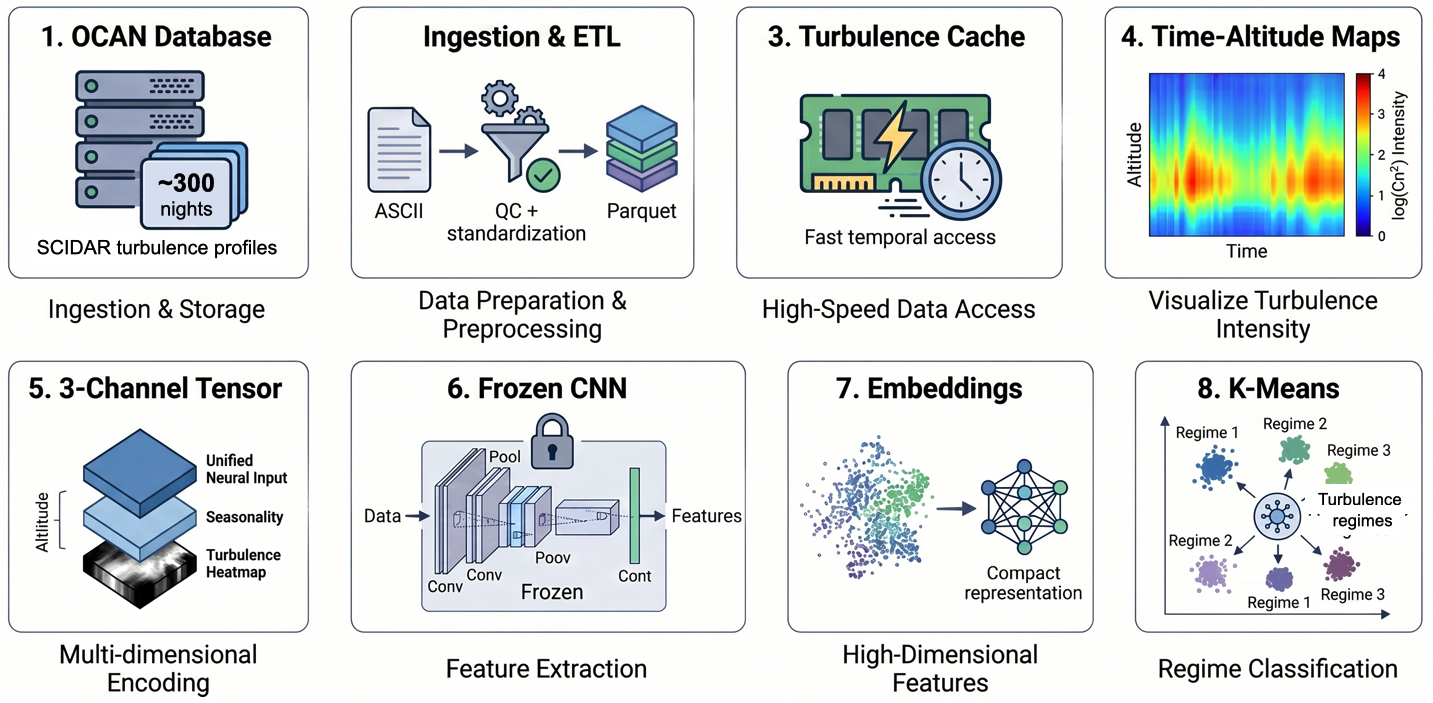}
   \end{tabular}
   \end{center}
   \caption[ ] 
   { \label{fig:infografia} Workflow of the CNN-based reanalysis pipeline. Generalized SCIDAR turbulence profiles are transformed through ETL processing, temporal caching, and temporal--altitude map generation into three-channel tensors suitable for CNN processing. Deep embeddings extracted from pre-trained networks are subsequently clustered to identify recurrent atmospheric turbulence regimes within the OCAN database.
} 
   \end{figure}

\subsection{ETL Architecture}

The OCAN turbulence database is organized as a collection of ASCII files, each containing all Generalized SCIDAR turbulence profiles acquired during that night. While convenient for data acquisition and archival purposes, this organization is not directly suited to the large-scale processing and repeated data access required by the analysis presented in this work. To overcome these limitations, we developed an Extract–Transform–Load (ETL) pipeline designed to ingest, standardize, and organize the full dataset within a unified analytical framework.

Each file contains multiple C$_n^2$(h) profiles acquired throughout the observing night, together with the metadata required to describe the corresponding measurements. During the extraction phase, all profile records and associated metadata were parsed and converted into tabular representations. The resulting datasets were then stored in columnar \texttt{Parquet} format, providing a compact and efficient storage format for the subsequent transformation and analysis stages.

The transformation stage focused on consolidating the measurements from both observatories into a homogeneous data structure suitable for large-scale analysis. Individual datasets were merged into site-level master DataFrames, from which redundant fields and duplicated records were removed following a series of consistency checks. To facilitate profile tracking throughout the processing chain, a unique identifier was assigned to each turbulence profile based on its observational metadata and acquisition timestamps. Additional temporal descriptors were also generated, providing a common framework for subsequent analyses of turbulence variability across different timescales and seasons.

A final preprocessing stage was applied to ensure the consistency of the turbulence measurements before the generation of derived data products. Non-physical turbulence values were removed, C$_n^2$ measurements were converted to a logarithmic scale, and profile completeness was verified across the vertical sampling range. To guarantee a common vertical extent across all measurements, all profiles were restricted to a maximum altitude of 20 km above sea level, while profiles with incomplete vertical coverage were discarded. The resulting datasets provide a homogeneous and quality-controlled representation of the OCAN turbulence archive, forming the basis for the temporal mapping and feature-extraction procedures described in the following sections.

\subsection{ Turbulence Cache}

Many of the analyses performed in this work require repeated access to large numbers of individual C$_n^2$(h) profiles distributed throughout the observational archive. Performing these operations directly on the processed site-level DataFrames proved computationally inefficient, particularly when exploring multiple temporal configurations involving different time windows and aggregation intervals.

To reduce the computational overhead associated with repeated profile retrieval, an intermediate turbulence cache was developed as an additional processing layer on top of the ETL products. The cache reorganizes the turbulence measurements according to their acquisition time, enabling rapid temporal queries without requiring repeated scans of the full datasets.

For each observatory, the processed turbulence profiles were grouped at one-second resolution and stored in a dictionary-like structure indexed by timestamp. Each entry contains the corresponding vertical turbulence profile represented on the common altitude grid established during the preprocessing stage. This organization allows direct retrieval of profiles associated with arbitrary temporal intervals while preserving the original temporal sampling of the observations.

The cache was designed to support multiprocessing workflows and repeated experimentation with different temporal aggregation schemes. By decoupling profile access from the original tabular datasets, it significantly reduces the computational cost of profile selection, temporal aggregation, and the generation of derived turbulence representations. The resulting cache constitutes the primary data source used throughout the subsequent stages of the analysis.

\subsection{Temporal--Vertical Turbulence Maps}

The turbulence cache described in the previous section enables the retrieval of large numbers of C$_n^2$(h) profiles over arbitrary temporal intervals. To characterize the temporal evolution of atmospheric turbulence, individual profiles were aggregated into temporal--vertical turbulence maps. These maps provide a two-dimensional representation in which the vertical axis corresponds to altitude, the horizontal axis represents time, and each matrix element contains the turbulence intensity measured at a particular altitude and instant. In this representation, consecutive C$_n^2$(h) profiles are combined into a continuous time--altitude description of atmospheric turbulence, preserving both the structure of individual turbulent layers and their temporal evolution.

From the perspective of site characterization, these maps provide a compact description of turbulence variability across both altitude and time. At the same time, their regular matrix structure makes them naturally compatible with image-processing techniques. Because each matrix element can be interpreted as a pixel whose value encodes turbulence intensity, these representations are hereafter referred to as \emph{turbulence heatmaps} when discussing the ML stages of the analysis.

For a selected temporal interval, all available C$_n^2$(h) profiles were retrieved from the turbulence cache and projected onto the common altitude grid established during the preprocessing stage. The resulting time--altitude matrices constitute the fundamental data product used throughout this work and provide the image-like representations from which the deep-learning features are subsequently extracted.

\subsection{Heatmap Generation}

Temporal--vertical turbulence maps were generated by aggregating C$_n^2(h)$ profiles retrieved from the turbulence cache over predefined temporal intervals. For each map, the horizontal axis represents time, and the vertical axis corresponds to altitude on the common grid established during preprocessing.

Two parameters control the map construction: the heatmap timeframe, which defines the total temporal extent of the map, and the bin duration, which specifies the temporal interval used to average individual turbulence measurements. Empty bins were filled using time-weighted interpolation between the nearest available observations, while bins containing measurements retained values derived exclusively from real data.

For each generated map, metadata describing temporal coverage, profile density, and interpolation statistics were computed and stored for subsequent quality assessment and ML processing.

\subsection{High-Quality Tensor Generation}

To ensure the reliability of the ML analysis, only heatmaps satisfying predefined quality criteria were retained. The selection considered temporal coverage, minimum profile density, and the maximum length of interpolated intervals.

Each accepted heatmap was converted into a three-channel tensor. The first channel contains the normalized turbulence intensity, while the remaining two channels encode the seasonal phase through sine and cosine transformations of the day of the year. This representation preserves compatibility with standard CNN architectures while incorporating temporal context into the input data.

\subsection{CNN Feature Extraction and Clustering}
\label{sec:cluster}

The feature-extraction stage was implemented using transfer learning ([\citenum{pan2010transfer}]). Pre-trained CNNs from the EfficientNet family [\citenum{tan2019efficientnet}] were employed as frozen feature extractors, with their final classification layers removed and only the convolutional backbone retained. This approach leverages feature representations learned from large-scale image datasets while avoiding the computational cost and data requirements associated with training deep networks from scratch.

Each three-channel turbulence tensor was processed by the CNN to obtain a compact embedding describing the spatial and temporal structure of the corresponding turbulence heatmap. Before feature extraction, tensors were padded when necessary in order to preserve their original aspect ratio and temporal--vertical structure, avoiding the distortions that could be introduced by image resizing. Rather than performing image classification, the networks were used exclusively to generate deep feature representations suitable for subsequent unsupervised analysis.

The resulting embeddings were clustered using the K-Means algorithm ([\citenum{macqueen1967kmeans}]). The optimal number of clusters was evaluated through inertia and silhouette analyses, while cluster quality was further assessed using standard validation metrics, including the Davies--Bouldin and Calinski--Harabasz indices. Low-dimensional projections obtained with PCA\footnote{PCA — Principal Component Analysis}, t-SNE\footnote{t-SNE — t-distributed Stochastic Neighbor Embedding} [\citenum{maaten2008tsne}], and UMAP\footnote{UMAP — Uniform Manifold Approximation and Projection} [\citenum{mcinnes2018umap}] were used to visualize the embedding space and support the interpretation of the resulting atmospheric regimes.

   \begin{figure} [ht]
   \begin{center}
   \begin{tabular}{c} 
   \includegraphics[width=12cm]{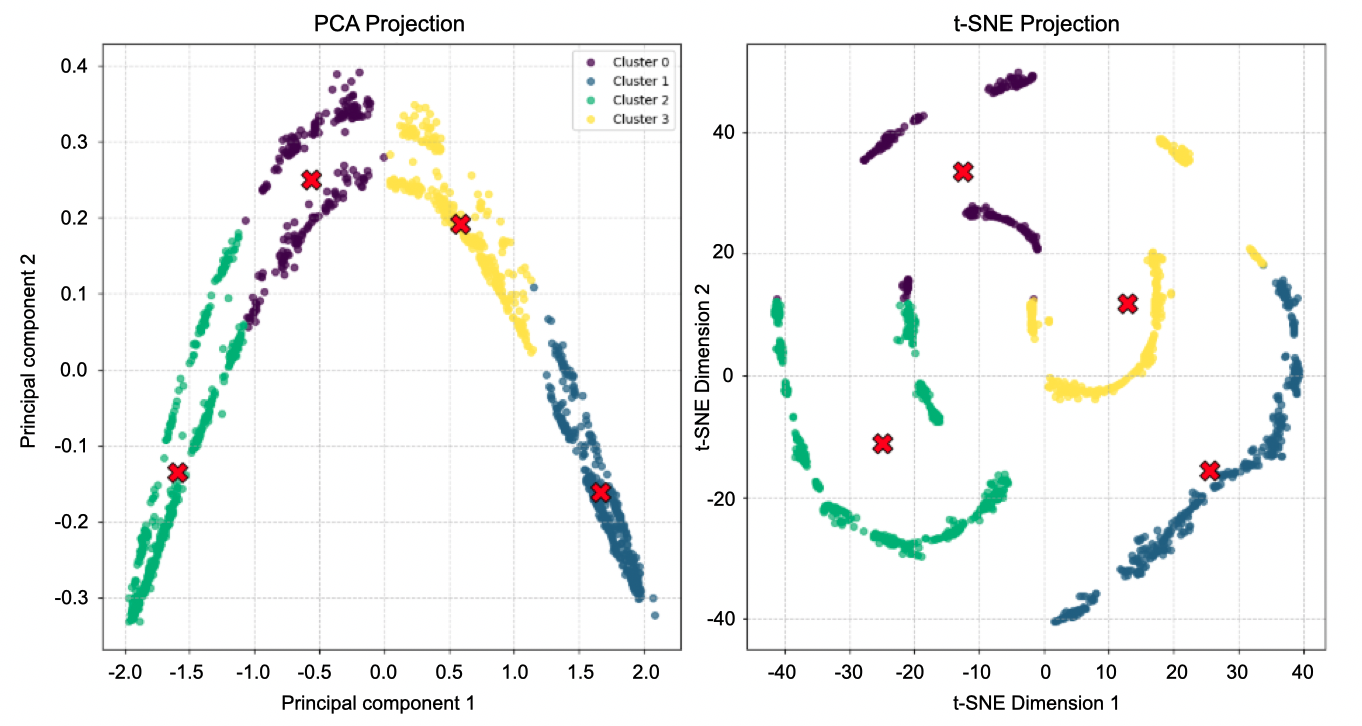} 
   \end{tabular}
   \end{center}
   \caption[ ] 
   { \label{fig:cluster_heatmaps} 
Cluster visualization for the full C$_n^2$(h) database for OCAN (k=4) using EfficientNetB1 and 180 seconds bins within 60 minutes timeframe.
}

   \end{figure} 

\section{RESULTS}


 The results presented below constitute a preliminary assessment of the proposed CNN-based analysis framework. The primary objective is to evaluate whether the methodology can identify coherent patterns within the turbulence database; therefore, the physical interpretation and climatological significance of the resulting clusters remain subjects of ongoing investigation.
 
 Based on the clustering validation metrics described in Section \ref{sec:cluster}, a four-cluster solution was adopted for the subsequent analysis. Figure~\ref{fig:cluster_heatmaps} presents the mean temporal--vertical turbulence maps corresponding to the four turbulence regimes identified from the complete OCAN database.

   \begin{figure} [ht]
   \begin{center}
   \begin{tabular}{c} 
   \includegraphics[width=8.25cm]{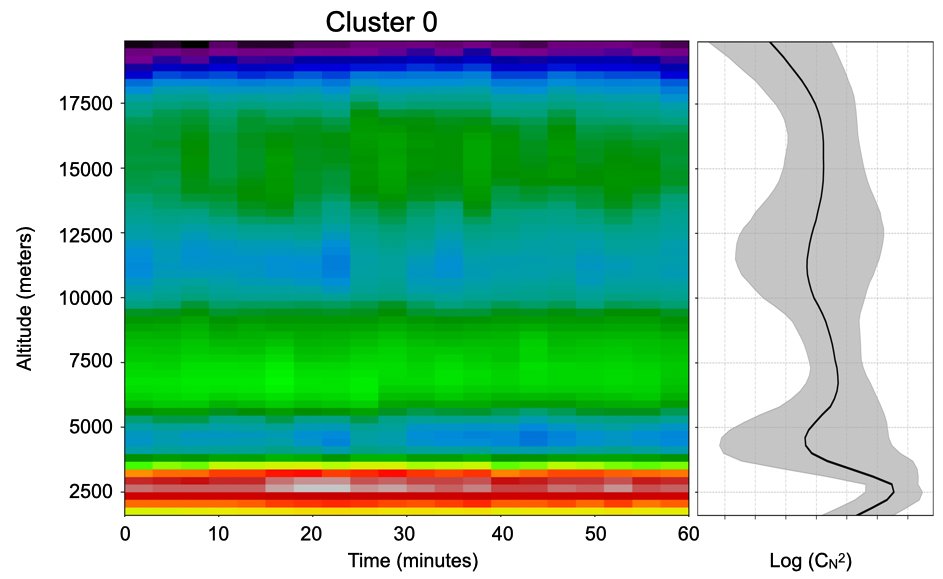} 
   \includegraphics[width=8.25cm]{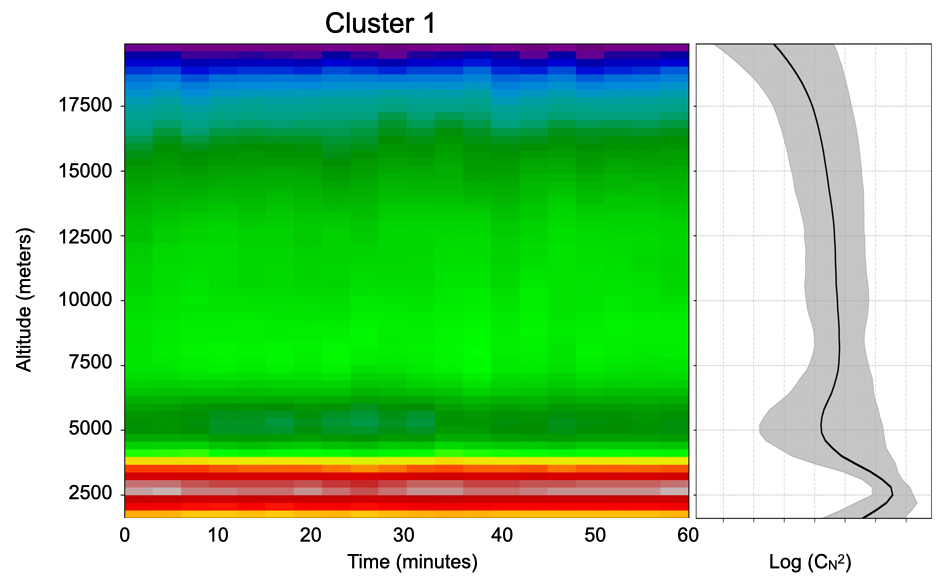} \\
   \includegraphics[width=8.25cm]{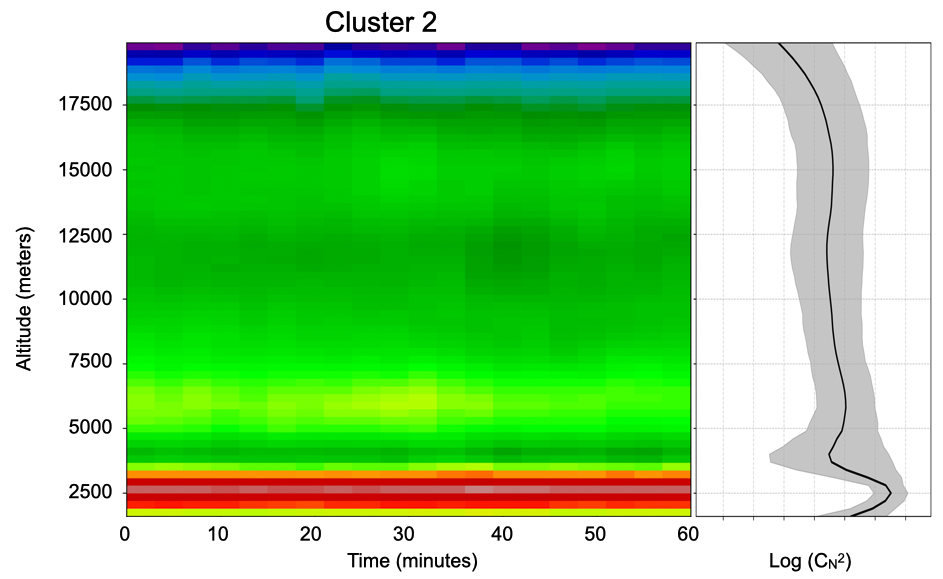} 
   \includegraphics[width=8.25cm]{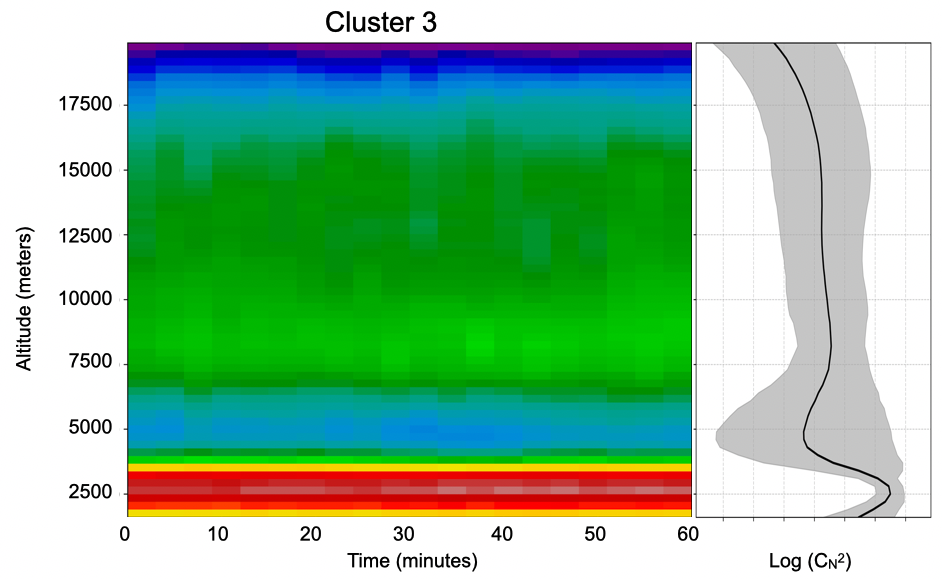} \\
   \end{tabular}
   \end{center}
   \caption[ ] 
   { \label{fig:cluster_heatmaps} 
Mean temporal--vertical turbulence maps associated with the four turbulence regimes identified from the complete OCAN database. Each panel represents the average heatmap obtained from all samples assigned to a given cluster. The horizontal axis corresponds to time within the 60-minute heatmap window, while the vertical axis represents altitude above sea level. The colour scale represents the average turbulence intensity derived from logarithmic C$_n^2$ values and expressed in normalized arbitrary units, with warm colours (yellow to red) indicating stronger turbulence, cool colours (blue to purple) indicating weaker turbulence, and green tones representing intermediate turbulence levels. The four regimes show distinct vertical turbulence structures, particularly in the relative strength and altitude distribution of turbulence layers. These differences are also clearly reflected in the mean vertical turbulence profiles shown in the right-hand panels, which reveal cluster-dependent variations in the altitude and intensity of the dominant turbulence layers.
}

   \end{figure} 
 
 The resulting regimes suggest the presence of distinct vertical turbulence structures, indicating that the CNN embeddings preserve physically meaningful information from the original C$_n^2$(h) profiles. Although all clusters are characterized by strong turbulence near the observatory level, clear differences are observed in the intensity and altitude distribution of turbulence layers.

 Figure \ref{fig:seasonal} presents the seasonal distribution of these regimes. A clear
seasonal organization is observed, with each cluster dominating during specific periods of the year. Cluster 3
is primarily associated with the winter months and reappears during late spring and early summer, whereas
Cluster 1 dominates most of the spring season. Cluster 0 is preferentially observed during summer and early
winter, while Cluster 2 becomes the predominant regime from late summer through autumn. The relatively sharp
transitions between dominant clusters suggest that the identified regimes are linked to recurrent atmospheric
conditions rather than representing a continuous spectrum of turbulence states. In contrast to traditional analyses that examine turbulence statistics within predefined calendar seasons (e.g. [\citenum{2009Garcia-Lorenzo}]), the present clustering approach identifies seasonal patterns that emerge naturally from the underlying turbulence structure.

 While these differences suggest that the CNN embeddings preserve meaningful information from the original turbulence profiles, additional validation is required to determine whether the identified clusters correspond to physically distinct atmospheric states.
 
To assess the physical significance of the identified turbulence regimes, wind-roses distributions associated with each cluster were analysed. Wind roses were generated separately for ORM (Fig. \ref{fig:windorm}) and OT (Fig. \ref{fig:windot}), since the two observatories are subject to markedly different local orographic conditions. Wind data at ORM and OT come from the automatic weather stations of the telescopes JKT \footnote{Jacobus Kapteyn Telescope. Latitude  28N45'40.1'', longitude: 17W52'40.9'', altitude: 2374~m.a.s.l.The station is installed on a rooftop, with the anemometer mounted on a 12-m mast located 8 m from the SCIDAR telescope and less than 5 m from the telescope dome in the prevailing leeward direction.} and GONG\footnote{Global Oscillation Network Group. Latitude  28N18'2.4'', longitude 16W30'43.8'', altitude: 2395~m.a.s.l. The station is installed on a rooftop, with the anemometer mounted on a 7-m mast located 124 m from the SCIDAR telescope.}, respectively. Both stations are placed in small masts on roofs; JKT's is less than 5~m close to the telescope spherical dome to the typical leeward direction. Despite the overall similarities in their large-scale turbulence behaviour, the clusters exhibit distinct wind-direction signatures at both sites. This suggests that the CNN-derived regimes are linked to different atmospheric circulation patterns rather than representing purely statistical groupings. Because wind information was not used during the clustering process, this result provides an independent validation of the physical relevance of the identified turbulence classes.

   \begin{figure} [ht]
   \begin{center}
   \begin{tabular}{c} 
   \includegraphics[width=16cm]{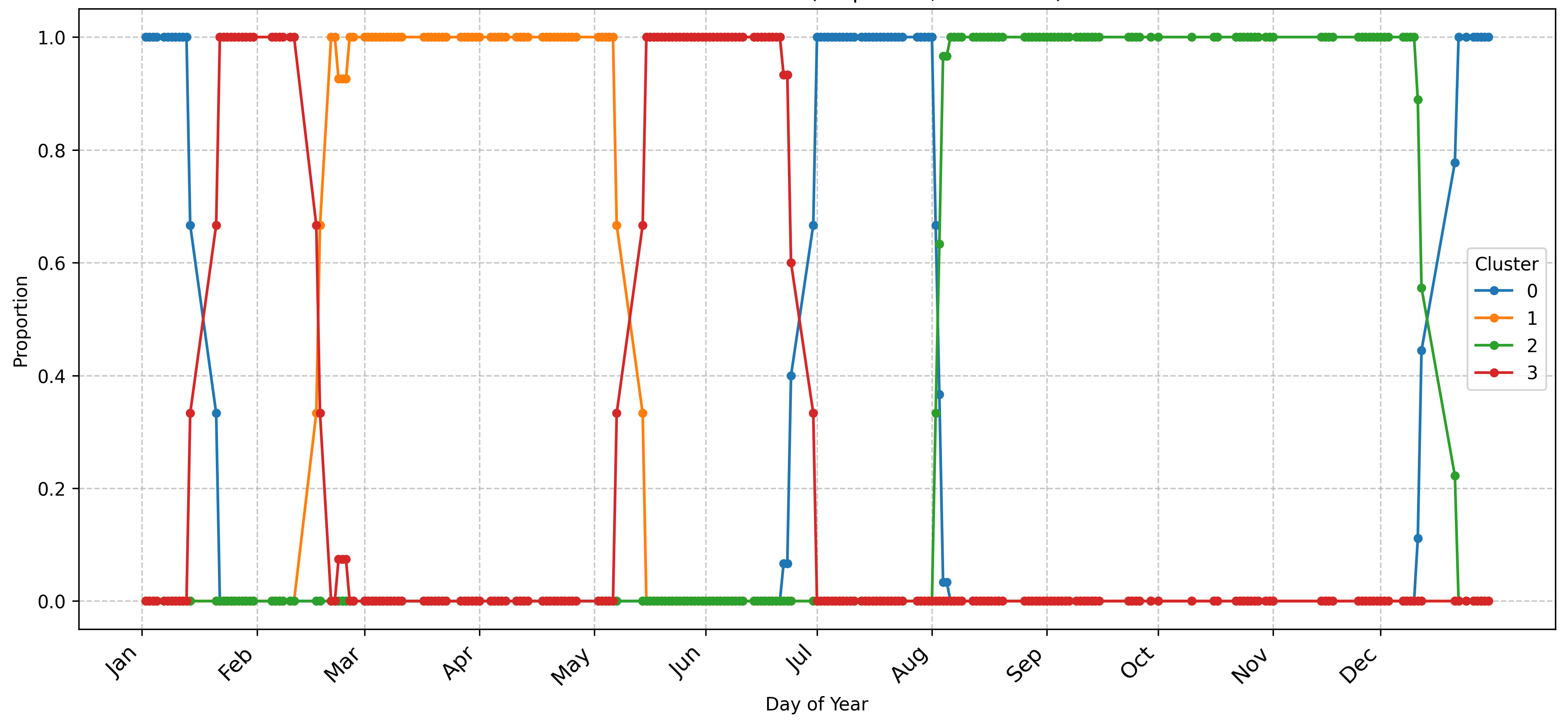} \\
   \end{tabular}
   \end{center}
   \caption[ ] 
   { \label{fig:seasonal} Seasonal distribution of the four turbulence regimes identified through CNN-based clustering of the complete OCAN dataset. The curves show the relative proportion of heatmaps assigned to Cluster 0 (blue), Cluster 1 (orange), Cluster 2 (green), and Cluster 3 (red) as a function of the day of the year. The analysis includes all 238 high-quality turbulence heatmaps, and the distributions were smoothed using a 3-day moving window. Distinct seasonal preferences are observed among the identified regimes, with different clusters dominating during specific periods of the year.
} 
   \end{figure} 

   \begin{figure} [ht]
   \begin{center}
   \begin{tabular}{c} 
   \includegraphics[width=14cm]{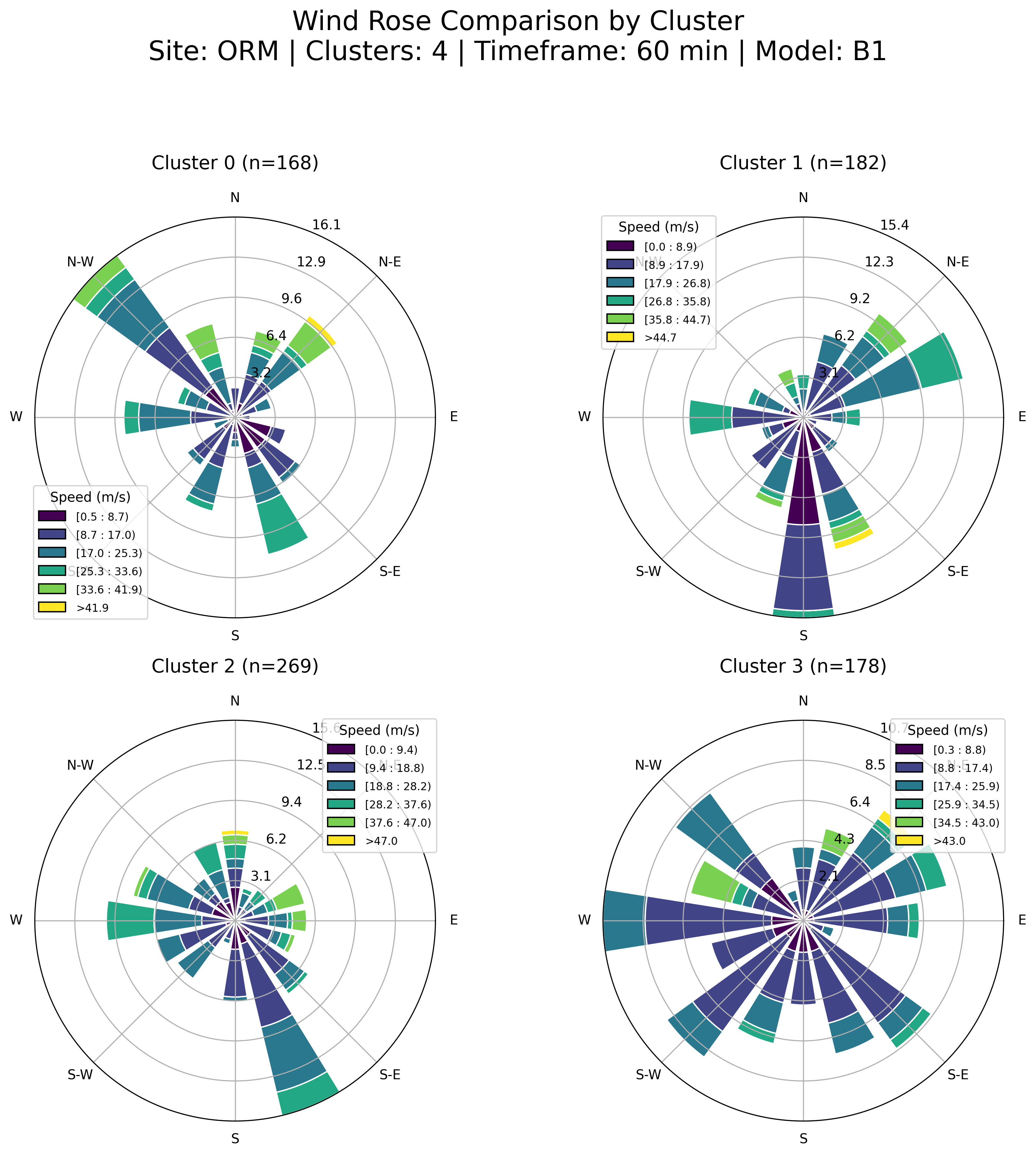} 
   \end{tabular}
   \end{center}
   \caption[ ] 
  { \label{fig:windorm} Wind-rose distributions associated with the four turbulence regimes identified at the ORM. Each panel shows the distribution of wind direction and wind speed for all heatmaps assigned to a given cluster. The clusters exhibit distinct prevailing wind patterns and speed distributions, indicating that the CNN-derived turbulence regimes are associated with different atmospheric circulation conditions. Since wind information was not used during the feature-extraction or clustering stages, these differences suggest the physical relevance of the identified turbulence classes.
} 
   \end{figure} 

   \begin{figure} [ht]
   \begin{center}
   \begin{tabular}{c} 
   \includegraphics[width=14cm]{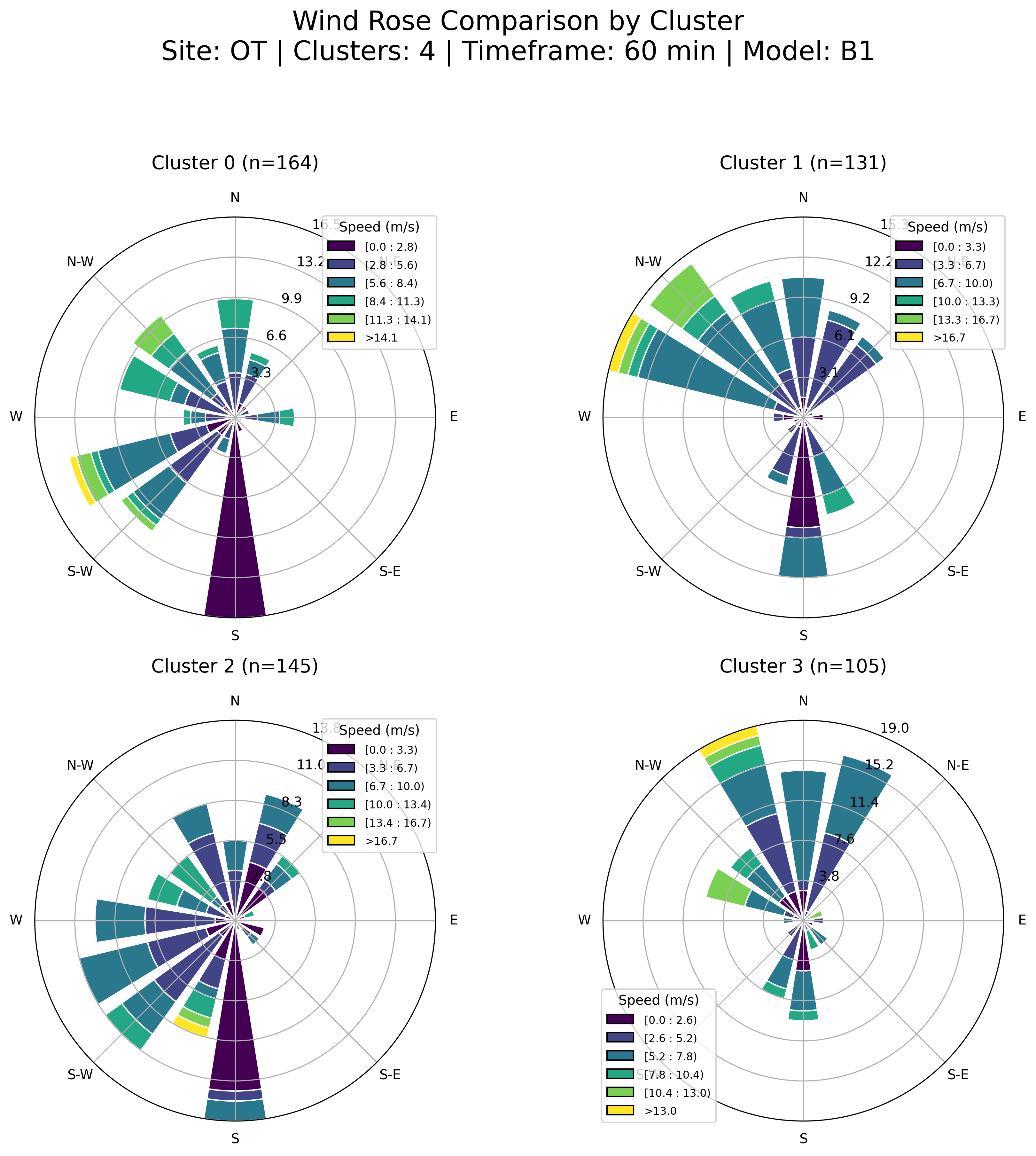} 
   \end{tabular}
   \end{center}
   \caption[ ] 
  { \label{fig:windot} Same as Fig. \ref{fig:windorm} but for OT.

} 
   \end{figure} 

\section{Conclusions}

In this work, we presented a CNN-based reanalysis of the complete OCAN atmospheric turbulence database, comprising nearly 300 observing nights and hundreds of thousands of Generalized SCIDAR turbulence profiles acquired at the ORM and the OT.

A processing framework was developed to transform the original C$_n^2$(h) measurements into temporal--vertical turbulence representations suitable for ML analysis. By combining turbulence heatmaps, seasonality-aware tensor representations, transfer-learning feature extraction, and unsupervised clustering, it was possible to identify recurrent turbulence regimes directly from the full vertical and temporal structure of the measurements.

The resulting clusters exhibit distinct turbulence-layer distributions and clear differences in their associated surface wind-roses patterns at both observatories. Because wind information was not included in the CNN embeddings or clustering procedure, the observed wind-regime signatures suggest that the identified turbulence classes capture physically meaningful differences in the atmospheric conditions. These results suggest that the extracted embeddings preserve physically meaningful information from the original turbulence profiles and are capable of separating recurrent atmospheric circulation regimes.

The present study should be regarded as a preliminary exploration of the applicability of deep-learning techniques to large optical-turbulence archives. Future work will focus on a more detailed physical interpretation of the identified regimes, including their relationship with adaptive-optics parameters, meteorological conditions, and large-scale atmospheric circulation patterns. Preliminary tests performed with multiple CNN architectures and heatmap temporal configurations suggest that the quality of the extracted embeddings is strongly influenced by the correspondence between the heatmap dimensions and the characteristic spatial scales captured by the network. The best-performing models were not necessarily those with the highest architectural complexity, but those whose convolutional structure was better matched to the temporal–vertical resolution of the turbulence maps. This result highlights the importance of jointly optimizing the heatmap generation parameters and the CNN architecture when applying transfer-learning techniques to optical-turbulence characterization.

Overall, the results indicate that CNN-based feature extraction combined with unsupervised learning constitutes a promising framework for the characterization and exploration of long-term atmospheric turbulence databases. Beyond identifying distinct turbulence structures, the methodology reveals links between turbulence variability and large-scale atmospheric circulation patterns, opening new possibilities for data-driven site characterization and adaptive-optics studies.

\acknowledgments 
 
 The authors acknowledge support from the Spanish Ministry of Science and Innovation through the Spanish State Research Agency (AEI-MCINN/10.13039/501100011033) via the grants “Participation of the Instituto de Astrofísica de Canarias in the development of HARMONI: Delta-D1 phase and Rescope Study” (PID2022-140483NB-C21 and PID2024-158231NB-C21). The authors also acknowledge support from the project “IPOICE: Apoyo al desarrollo en el IAC de la preóptica (IPO) y el control electrónico (ICE) de HARMONI”, funded by ESO-MICIN and the European Union NextGenerationEU/RTRP (Agreement No. 64365/ESO/15/66976/JSC). The authors acknowledge the use of ChatGPT (OpenAI) for assistance with language editing and improving the clarity of the manuscript.

\bibliography{report} 
\bibliographystyle{spiebib} 

\end{document}